\documentclass[11pt]{cernrep}
\usepackage{graphicx}
\usepackage{here}
\usepackage{epsf}
\usepackage{epsfig}
\usepackage{subfigure}
\usepackage{rotate}

\renewcommand{\thesection}{\arabic{section}}
\renewcommand{\thesubsection}{\arabic{section}.\arabic{subsection}}

\def \be {\begin{equation}}
\def \ba {\begin{eqnarray}}
\def \ee {\end{equation}}
\def \ea {\end{eqnarray}}

\begin{document}

\pagestyle{empty}
\begin{titlepage}
\null
\vspace{2cm}
\begin{center}
\Large\bf A Note on the $J/\psi$ 
Strong Couplings{\footnote{in 
Workshop on Hard Probes in Heavy Ion Collisions at the LHC}}
\end{center}
\vspace{1.5cm}

\begin{center}
\begin{large}
A. Deandrea\\
\end{large}
\vspace{0.5cm}
Institut de Physique Nucl\'eaire, Universit\'e de Lyon I\\  
4 rue E.~Fermi, F-69622 Villeurbanne Cedex, France\\
\vspace{0.7cm}
\begin{large}
G. Nardulli \\
\end{large}
\vspace{0.5cm}
Dipartimento di Fisica, Universit\`a di Bari and INFN Bari,\\
via Amendola 173, I-70126 Bari, Italia\\
\vspace{0.7cm}
\begin{large}
A. D. Polosa\\
\end{large}
\vspace{0.5cm}
CERN - Theory Division \\
CH-1211 Geneva 23, Switzerland\\ 

\vspace{0.5cm}

\end{center}

\vspace{1.3cm}

\begin{center}
\begin{large}

{\bf Abstract}\\[0.5cm]
\end{large}
\parbox{14cm}{In this note we present an evaluation of the couplings
$JD^{(*)}D^{(*)}$ and $JD^{(*)}D^{(*)}\pi$ in the Constituent
Quark Model. These couplings are a crucial ingredient in the
calculation of cross sections for the processes $\pi J/\psi\to
D^{(*)}\bar{D}^{(*)}$, an  important background for the $J/\psi$
suppression signal in quark-gluon plasma.}
\end{center}

\vspace{1.5cm}
\noindent
BARI-TH/02-450\\
CERN-TH/2002-343 \\
LYCEN-2002-68\\
November 2002
\end{titlepage}

\title{A NOTE ON THE $J/\psi$ 
STRONG COUPLINGS\\[-2.3cm]}
\author{{A. Deandrea, G. Nardulli and A.D. Polosa}}
\maketitle

\begin{flushright}
~\\[-8cm]CERN--TH/2000--101\\[6.6cm]
\end{flushright}

\begin{abstract}
In this note we present an evaluation of the couplings
$JD^{(*)}D^{(*)}$ and $JD^{(*)}D^{(*)}\pi$ in the Constituent
Quark Model. These couplings are a crucial ingredient in the
calculation of cross sections for the processes $\pi J/\psi\to
D^{(*)}\bar{D}^{(*)}$, an  important background for the $J/\psi$
suppression signal in quark-gluon plasma.
\end{abstract}

This note is a preliminary report on a study of absorption effects of the 
$J/\psi$ resonance due to its interaction with the hot hadronic
medium formed in relativistic heavy ion scattering. We will give
the full analysis elsewhere \cite{noi}; here we limit the
presentation to the study of the strong couplings of $J/\psi$, low
mass charmed mesons and pions. In the calculation of the relevant
cross sections one encounters tree-level diagrams such as those
depicted in  Fig.~1. Previous studies of these effects can be
found in \cite{cina1}-\cite{cina3}. Besides the $g(DD^*\pi)$
couplings, for which  both theoretical \cite{theor},\cite{report}
and experimental \cite{exp} results are available, in Fig.~1 the
$JD^{(*)}D^{(*)}$ and $JD^{(*)}D^{(*)}\pi$ couplings appear. They
have been estimated by different methods, that are, in our
opinion, unsatisfactory. For example the use of the
$SU(4)$ symmetry puts on the same footing the heavy quark $c$ and
the light quarks, which is at odds with the results obtained
within the Heavy Quark Effective Theory (HQET), where the opposite
approximation $m_c\gg \Lambda_{QCD}$ is used (for a short review
of HQET see \cite{report}). Similarly, the rather common approach
based on the Vector Meson Dominance (VMD) should be considered
critically, given the large extrapolation $p^2=0\to m_{J/\psi}^2$
that is involved. A different evaluation, based on QCD Sum Rules can
be found in \cite{qcdsr} and presents the typical theoretical
uncertainties of this method. In this note we will use 
another approach, based on the Constituent Quark Model (CQM),
which is a quark-meson model taking into account explicitly the
HQET symmetries (for more details on the CQM see \cite{nc}).

The CQM model has turned out to be particularly suitable for the
study of exclusive heavy meson decays. Since its Lagrangian
contains the Feynman rules for vertices formed by a heavy
meson, a heavy quark and a light quark, transition amplitudes are
computable via simple (constituent) quark loop diagrams where
mesons enter as external legs. The model is relativistic and
incorporates, besides  the heavy quark symmetries, also the chiral
$SU_2$ symmetry of the light quark sector.
 The calculation of the
$g(DD^*\pi)$ coupling constant in the CQM can be found in
\cite{nc}. Here we will consider the calculation of the
$JD^{(*)}D^{(*)}$ and $JD^{(*)}D^{(*)}\pi $ vertices and, to begin
with, the $JD^{(*)}D^{(*)}$ vertex, whose
calculation proceeds via a VMD ansatz (see Fig. 2).

In the CQM the evaluation of the loop diagram depicted on the
l.h.s. of the VMD equation in Fig.~2  amounts to the calculation
of the Isgur-Wise function, which can be found in \cite{nc}. The
result is \be \xi(\omega)=Z_H\left[ \frac{2}{1+\omega}
I_3(\Delta_H)+
\left(m+\frac{2\Delta_H}{1+\omega}I_5(\Delta_H,\Delta_H,\omega)\right)\right]\
, \ee where the $I_i$ integrals are listed in the appendix. This
result arises from the calculation of the following loop integral
(for the $JDD$ process): \be m_D Z_H\frac{iN_c}{16\pi^4}\int
d^4\ell \frac{{\rm Tr}\left[(\gamma\cdot\ell+m) \gamma_5
(1+\gamma\cdot v^\prime)\gamma_\mu (1+\gamma\cdot
v)\gamma_5\right]}{4 (\ell^2-m^2)(v\cdot
\ell+\Delta_H)(v^\prime\cdot\ell-\Delta_H)}\ , \ee where \be
\frac{1+\gamma\cdot v}{2}\frac{1}{v\cdot k} \ee is the heavy quark
propagator of  the HQET, $v$ and $v^\prime$ are the $4-$velocities
of the two heavy quarks; they are assumed equal, in the infinite
quark mass limit, to the hadron $D^{(*)}$ velocities. On the other
hand $\omega=v\cdot v^\prime$.

 Let us also introduce $k$, the meson
residual momentum, defined by $p^\mu_D=m_c v^\mu + k^\mu$; it
enters the calculation through the parameter $\Delta_H=v\cdot k$
which is equal to the mass difference $m_D-m_c$. Its numerical
value is  in the range $0.3-0.5$~GeV \cite{nc}.  If we consider a
$D^*$ meson instead of a $D$, a factor $-\gamma_5$ must be
substituted by $\gamma\cdot\epsilon$, $\epsilon$ being the
polarization of $D^*$. The constant $Z_H$ arises from
 the coupling of $D^{(*)}$ mesons
to their constituent  quarks (more precisely the coupling constant
is $\sqrt{Z_H m_{D}}$); $Z_H$ is computed and tabulated
in~\cite{nc}.

 We note that  the Isgur-Wise function obeys the
normalization condition $\xi(1)=1$, arising from  the flavor
symmetry of the HQET. This is the Luke's theorem, whose ancestor
for the light flavors is the Ademollo-Gatto theorem~\cite{agl}.
 The explicit definition of the Isgur-Wise form factor is:
\be \langle H(v^\prime)|\bar c \gamma_\mu c
|H(v)\rangle=-\xi(\omega){\rm Tr}\left(\bar H\gamma_\mu H\right)\
. \ee Here $H$ is the multiplet containing both the $D$ and the
$D^*$ mesons   \cite{report}: \be H=\frac{1+\gamma\cdot
v}{2}(-P_5\gamma_5+ \gamma\cdot P)\ ,\ee and $P_5,\,P^\mu$ are
annihilation operators for the charmed mesons. As an example,
for the transition between two pseudoscalar mesons $D$ one finds:
\be \langle D(v^\prime)|\bar c \gamma_\mu c
|D(v)\rangle=m_D\xi(\omega)(v+v^\prime)_\mu. \ee
 One can compute in the CQM
the Isgur-Wise function for any value of $\omega$ and not only in
the region $\omega>1$, which is experimentally accessible $via$
the semileptonic $B\to D^{(*)}$ decays. Since \be
\omega=\frac{p_1^2+p_2^2-p^2}{2 \sqrt{p_1^2p_2^2}}\ee
($p_1,\,p_2=$ momenta of the two $D$ resonances), differently from
the naive use of VMD, by our method we can have a control on the
dependence on $p^2$ (and also on the off-shell behavior in the
variables $p_1^2,\,p_2^2$).

Let us now consider the r.h.s. of the equation depicted in Fig. 2.
For the coupling of $J/\psi$ to the current we use the matrix
element
\be \langle 0|\bar c \gamma^\mu c
|J(q,\eta)\rangle=f_Jm_{J/\psi}\epsilon^\mu \ee with $f_J=0.405\pm
0.014$ GeV. As to the strong couplings $JD^{(*)}D^{(*)}$, the model
in Fig. 2 gives  the following effective lagrangians \ba {\cal
L}_{JDD}&=&ig_{JDD}\left(\bar{D}
{\stackrel{\leftrightarrow}{\partial}}_{\nu}D\right)J^\nu\ , \cr
{\cal L}_{JDD^*}&=&ig_{JDD^*}\epsilon^{\mu\nu\alpha\beta}J_{\mu}
\partial_{\nu}\bar D\partial_{\beta}D^*_\alpha\ ,\cr&&\cr
{\cal L}_{JD^*D^*}&=& ig_{JD^*D^*}\Big[ \bar{D}^{*\mu}\left(
{\partial}_{\mu}D^*_\nu\right)J^\nu - {D}^{*\mu} \left(
{\partial}_{\mu}\bar D^*_\nu\right)J^\nu \cr &-& \left(
 \bar{D}^{*\mu}{\stackrel{\leftrightarrow}{\partial}}_{\nu}
D^*_\mu\right)J^\nu \Big]\ .
\ea

As a consequence of the spin symmetry of the HQET we find: \ba
g_{JD^*D^*}&=&g_{JDD}\ ,\cr &&\cr
g_{JDD^*}&=&\frac{g_{JDD}}{m_D}\
,\ea while the VMD ansatz gives:
 \be \label{eq:coupling1}
g_{JDD}(p_1^2,\,p_2^2,\,p^2)=\frac{m^2_{J/\psi}-p^2}{f_J
m_{J/\psi}}\xi(\omega) \ .\ee Since $g_{JDD}$ has no zeros, eq.
(\ref{eq:coupling1}) shows that $\xi$ has a pole at
$p^2=m^2_{J/\psi}$, which is what one expects on the basis of
dispersion relations arguments. The CQM evaluation of $\xi$ does
show a strong peak for $p^2\approx(2m_c)^2$, even though, due to
$\displaystyle {\cal O}\,\left(\frac{1}{m_c}\right)$ effects, the
location of the singularity is not exactly at $p^2=m^2_{J/\psi}$.
This is shown in Fig. 3 where we plot
$g_{JDD}(p_1^2,\,p_2^2,\,p^2)$ for on shell $D$ mesons, as a
function of $p^2$ (the plot is obtained for $\Delta_H=0.4$~GeV and
$Z_H=2.36~$GeV$^{-1}$). For $p^2$ in the range $(0,4)$ GeV$^2$,
$g_{JDD}$ is almost flat, with a value \be g_{JDD}=8.0\pm 0.5\
.\label{gjdd}\ee For larger values of $p^2$ the method is
unreliable due to the above-mentioned incomplete cancellation
between the kinematical zero and the pole (the distorted shape
around the $J/\psi$ pole suggests that the contribution of the
nearby $\psi(2S)$ pole could also be relevant). Therefore, we
extrapolate the smooth behavior of $g_{JDD}$ in the small $p^2$
region up to $p^2=m^2_{J/\psi}$ and assume the validity of the
result (\ref{gjdd}) also for on-shell $J/\psi$ mesons. On the
other hand in the $p_1^2,p^2_2$ variables we find a smooth
behavior, compatible with that produced by a smooth form factor.
Let us finally observe that the result (\ref{gjdd}) agrees with
the outcome of the QCD sum rule analysis of \cite{qcdsr}; the
smooth behavior of the form factor found in \cite{qcdsr} agrees
with our result. This is not surprising, as the QCD sum rules 
calculation involves a perturbative part and a non perturbative 
contribution which is however suppressed; the perturbative term has its
counterpart in CQM in the loop calculation of Fig. 2 and the
overall normalization should agree as a consequence of the Luke's
theorem.

Let us now consider the $JD^{(*)}D^{(*)}\pi$ couplings. As
discussed  in \cite{falk}, but see also \cite{report}, the leading
contributions to the current matrix element $\langle
H(v^\prime)\pi |\bar c \gamma^\mu c|H(v)\rangle $ in the soft pion
limit (SPL) are the pole diagrams. The technical reason is that,
in the SPL, the reducing action of a pion derivative in the matrix
element is compensated in the polar diagrams by the effect of the
denominator that vanishes in the combined limit $q_\pi\to
0,\,m_c\to\infty$. Since the effect of the pole diagrams is
explicitly taken account in Fig. 1, we should not include any
further contribution. In any event, for the sake of a numerical
comparison, let us  consider the coupling $g_{JDD\pi}$; it can be
obtained  by a VMD ansatz similar to Fig. 2, but now the l.h.s is
modified by the insertion of a soft pion on the light quark line
(with a coupling $q_\pi^\mu/f_\pi \gamma_\mu\gamma_5$). We call
$\xi^\pi(\omega)$ the analogous form factor in the soft pion limit
and we find:
 \be \xi^\pi(\omega)=Z_H\left[
\frac{4m+2\Delta_H}{1+\omega}I_4(\Delta_H)-
\left(m^2+\frac{2\Delta_H^2+4m\Delta_H}{1+\omega}\right)
\frac{\partial I_5(\Delta_H,\Delta_H,\omega)}{\partial m^2}
\right] \ee (the integral $I_4$ is given in the appendix) . On
the other hand from the VMD ansatz of Fig. 2 we get \be {\cal
L}_{JDD\pi}=ig_{JDD\pi}\epsilon^{\mu\nu\alpha\beta}J_{\mu}
\partial_{\nu}D\partial_{\alpha}\bar{D}\partial_{\beta}\pi\ee and \be
\label{eq:coupling2}
g_{JDD\pi}(p_1^2,\,p_2^2,\,p^2)=\frac{(m_J^2-p^2)\xi^{\pi}
(\omega)}{f_\pi f_J m_D m_J}. \ee

In Fig.~4 we plot our result for  the $g_{JDD\pi}$ coupling with 
on shell $D$ mesons.
By the same arguments used to determine $g_{JDD}$ in Fig. 3 we
get, with all mesons on the mass-shell, \be
 g_{JDD\pi}=125\pm
15~{\rm GeV}^{-3}\ . \label{gddjp} \ee Let us now compare this
result with the effective $JDD\pi$ coupling obtained by a polar
diagram with an intermediate $D^*$ state. We get in this case \be
 g_{JDD\pi}^{polar}\approx\frac{g_{JDD^*}g_{D^*D\pi}}{2q_\pi\cdot p_D}
\ .\ee All the calculations presented in this note are valid in
the SPL, therefore one should consider  pion momenta not larger
than a few hundred MeV. Using \cite{exp} the result
$g_{D^*D\pi}=2m_D/f_\pi g $, with $g=0.59\pm 0.01\pm 0.07$, 
we get therefore
$g_{JDD\pi}^{polar}\approx 393, 196, 98$ GeV$^{-3}$ for $|q_\pi|$
respectively equal to $50,\,100,\,200$ MeV. This analysis shows
that, within the region of validity of the model,  in spite of the
rather large value of the coupling (\ref{gddjp}) the diagrams
containing this coupling are in general suppressed. 
Similar conclusions are
reached considering $D^*$ mesons instead of $D$ mesons.

Let us finally discuss the kinematical limits of our approach. To
allow the production of a $D^{(*)}D^{(*)}$ pair, as shown in Fig.
1, we must extend the region of validity of the model beyond the
SPL, since the threshold for the charmed meson pair is $|\vec
q_\pi|=700-1000~$MeV. The CQM, as the other models existing in the
literature, is a chiral model and this puts limits on the pion
momenta. Therefore one has to include a form factor enhancing the
small pion momenta region, for example \be f(|\vec
q_\pi|)=\frac{1}{\displaystyle 1+\frac{|\vec q_\pi|}{m_\chi}}\ .
\label{f}\ee A similar form factor is considered in \cite{cina1},
with a different motivation. Here we introduce it to ensure the
validity of our approach (in this sense the cross sections we can
compute by this model should be considered as a lower bound).
Since the main effect of (\ref{f})  should be that of reducing
contributions from pion momenta larger than a few hundred MeV, a
reasonable estimate for $m_\chi$ is in the range 400-600 MeV. This
choice implies that the direct couplings of Fig. 1 (diagrams 1c,
2d and 3e) should not dominate the final result since their
contribution is larger where the form factor is more effective.

\subsection*{Appendix}
We list the expressions used to numerically compute the integrals
$I_i$ quoted in the text. The ultraviolet cutoff $\Lambda$, the
infrared cutoff $\mu$ and the light constituent mass $m$ are fixed
in the model~\cite{nc} to be $\Lambda=1.25$~GeV, $\mu=0.3$~GeV and
$m=0.3$~GeV.

\ba I_3(\Delta) &=& - \frac{iN_c}{16\pi^4} \int^{\mathrm
{reg}} \frac{d^4k}{(k^2-m^2)(v\cdot k + \Delta +
i\epsilon)}\nonumber \\ &=&{N_c \over {16\,{{\pi }^{{3/2}}}}}
\int_{1/{{\Lambda}^2}}^{1/{{\mu }^2}} {ds \over {s^{3/2}}} \; e^{-
s( {m^2} - {{\Delta }^2} ) }\; \left( 1 + {\mathrm {erf}}
(\Delta\sqrt{s}) \right)\\
I_4(\Delta)&=&\frac{iN_c}{16\pi^4}\int^{\mathrm {reg}}
\frac{d^4k}{(k^2-m^2)^2 (v\cdot k + \Delta + i\epsilon)}
\nonumber\\ &=&\frac{N_c}{16\pi^{3/2}}
\int_{1/\Lambda^2}^{1/\mu^2} \frac{ds}{s^{1/2}} \;
e^{-s(m^2-\Delta^2)} \; [1+{\mathrm {erf}}(\Delta\sqrt{s})]\\
I_5(\Delta_1,\Delta_2,\omega) &= & \frac{iN_c}{16\pi^4}
\int^{\mathrm {reg}} \frac{d^4k}{(k^2-m^2)(v\cdot k + \Delta_1 +
i\epsilon ) (v'\cdot k + \Delta_2 + i\epsilon )} \nonumber \\
 & = & \int_{0}^{1} dx \frac{1}{1+2x^2 (1-\omega)+2x
(\omega-1)}\times\nonumber\\ &&\Big[
\frac{6}{16\pi^{3/2}}\int_{1/\Lambda^2}^{1/\mu^2} ds~\sigma \;
e^{-s(m^2-\sigma^2)} \; s^{-1/2}\; (1+ {\mathrm {erf}}
(\sigma\sqrt{s})) +\nonumber\\
&&\frac{6}{16\pi^2}\int_{1/\Lambda^2}^{1/\mu^2} ds \;
e^{-s\sigma^2}\; s^{-1}\Big], \ea where
\begin{equation}
\sigma(x,\Delta_1,\Delta_2,\omega)={{{\Delta_1}\,\left( 1 - x
\right)  + {\Delta_2}\,x}\over {{\sqrt{1 + 2\,\left(\omega -1
\right) \,x + 2\,\left(1-\omega\right) \,{x^2}}}}}.
\end{equation}

%%%%%%%%%%%%%%%%%%%%%%%%%%%%%%%%%%%%%%%%%%%%%%%%%
\begin{figure}[ht]
\epsfysize=10truecm \centerline{\epsffile{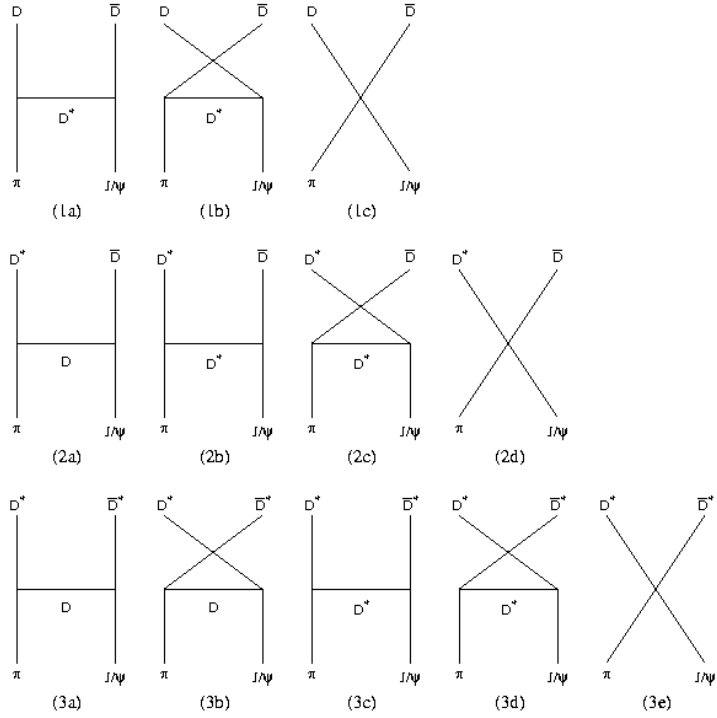}}
\caption{\label{fig} \footnotesize Feynman diagrams for $J/\psi$
absorption by the pion. (1) $J/\psi \pi \to D{\bar D}$, (2)
$J/\psi \pi \to {\bar D}D^*$ and $J/\psi \pi \to {\bar D}^* D^*$.}
\end{figure}

\begin{figure}[ht]
\begin{center}
\epsfig{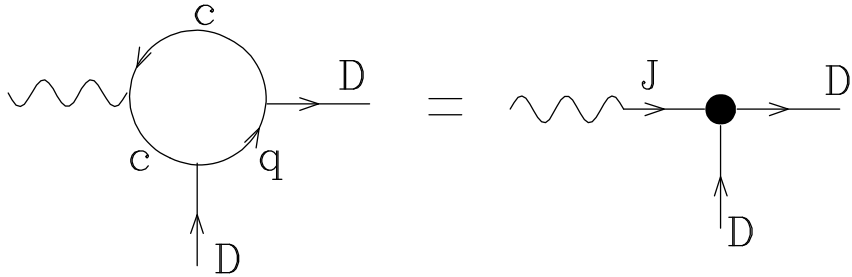}
\caption{\label{fig} \footnotesize The Vector Meson Dominance
equation giving the coupling of $J/\psi$ with $D,D^*$ in terms of
the Isgur-Wise function $\xi$. The function $\xi$ on the l.h.s. is
computed by a diagram with a quark loop. The coupling of each
$D^{(*)}$ meson to quarks is given by $\sqrt{Z_H m_D}$.}
\end{center}
\end{figure}

%\vskip2.0truecm

\begin{figure}[t!]
\begin{center}
\epsfig{height=4truecm,figure=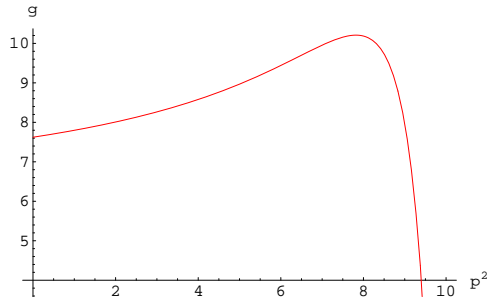}
\caption{\label{fig} \footnotesize The $p^2$ dependence of
$g=g_{JDD}(m_D^2,\,m^2_D,\,p^2)$, showing the almost complete
cancellation between the pole of the Isgur-Wise function and the
kinematical zero.  Units are  GeV$^2$ for $p^2$.}\end{center}
\end{figure}

\begin{figure}[t!]
\begin{center}
\epsfig{height=4truecm,figure=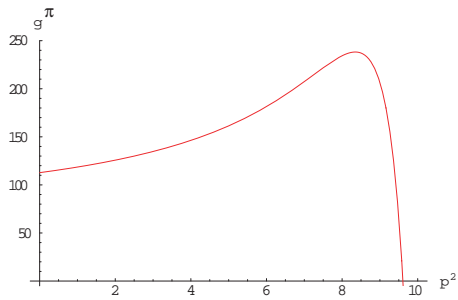}
\caption{\label{fig} \footnotesize The $p^2$ dependence of
$g^\pi=g_{JDD\pi}(m_D^2,\,m^2_D,\,p^2)$; as in Fig. 3 there is an almost
complete cancellation between the pole of the form factor and the
kinematical zero. Units are  GeV$^2$ for $p^2$ and GeV$^{-3}$ for
$g_{JDD\pi}$.}
\end{center}
\end{figure}

\end{document}